# Plasmonic Demultiplexer and Guiding

*Chenglong Zhao and Jiasen Zhang*[*]

State Key Laboratory for Mesoscopic Physics and Department of Physics, Peking

University, Beijing, 100871, China

Two-dimensional plasmonic demultiplexers for surface plasmon polaritons (SPPs), which consist of concentric grooves on a gold film, are proposed and experimentally demonstrated to realize light-SPP coupling, effective dispersion and multiple-channel SPP guiding. A resolution as high as 10 nm is obtained. The leakage radiation microscopy imaging shows that the SPPs of different wavelengths are focused and routed into different SPP strip waveguides. The plasmonic demultiplexer can thus serve as a wavelength division multiplexing element for integrated plasmonic circuit and also as a plasmonic spectroscopy or filter.

[*] E-mail: jszhang@pku.edu.cn

Recent developments on optics based on surface plasmon polaritons (SPPs) or plasmonics, which allow manipulating light in two dimensions with subwavelength scales,[1] have become a very active and attractive research field.[2,3] Taking advantage of the intrinsic two-dimensional (2D) nature of SPPs, which provides many opportunities for miniaturizing current optical devices to micro- and nano-size, various elements for the 2D SPP optics[4] analogous to the conventional 3D counterparts have been developed, such as lenses,[5-10] waveguides,[11,12] interferometers,[13,14] and microscopy.[15] By combining those plasmonic devices together, it may be possible to develop integrated plasmonic circuit and provide next-generation information network with improved bandwidth and speed.[16] Plasmonic demultiplexers, which are key elements of plasmonic circuits for wavelength division multiplexing (WDM) systems, need to be developed, and a high resolution is necessary to achieve a narrow channel spacing in high capacity plasmonic networks. Some SPP dispersing elements have been demonstrated, such as the overlapping bull's eye structure,[17] metal-insulator-metal resonator[18] and SPP crystal structures for splitting SPPs of different wavelengths into different directions.[19,20] However, a practical plasmonic demultiplexer needs to disperse multiple-channel data streams at different wavelengths spatially and focus them into various SPP wavelength components. In this work, we propose a plasmonic demultiplexer that can implement light-SPP coupling, effective dispersion, and multiple-channel SPP guiding. The resolution was as high as 10 nm in the experiment, and further improvement is expected by using a high-refractive index superstrate. Besides the applications in WDM systems, the plasmonic demultiplexer can also be used as a SPP spectroscopy or filter.

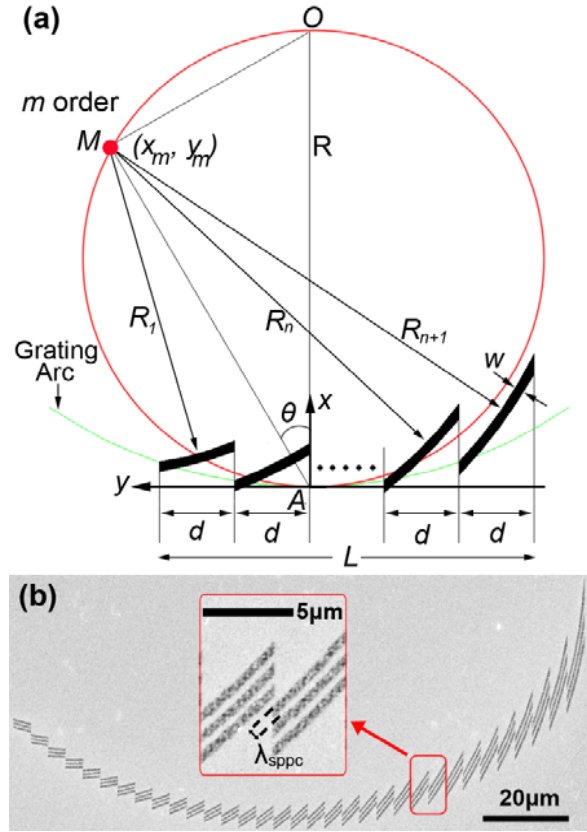

Figure 1. (a) Schematic of the plasmonic demultiplexer. (b) SEM image of a plasmonic demltiplexer with $m$ = 3. Inset shows the detail structures in the red retangluar box.

The plasmonic demultiplexer we proposed is schematically shown in Figure 1a, which is composed of concentric grooves on a gold film with the center $M$ and a radial width $w$. The projections of the grooves on the $y$ axis are spaced equidistantly and form a grating with a period $d$. If a plane wave illuminates the grooves normally, SPPs are launched at the grooves and diffracted according to the grating equation $dsin\theta = m\lambda_{SPP}$, where $m$ is the order number and an integer, $\theta$ is the diffracted angle, and $\lambda_{SPP}$ is the wavelength of the diffracted SPP. In order to focus the SPPs of different wavelengths on a focal circle (red circle in Figure 1a) with a diameter $R$ and touching the pole $A$ of a grating arc (green arc in Figure 1a) with a radius $R$, the grooves need to be located on the grating arc, which is a typical example of Rowland-type mounting.[21] To blaze the SPPs at different wavelength into intensive focal spots for the order number $m$, the concentric structures are designed, and the radial difference between the adjacent grooves, which determine the phase difference of the SPPs excited at the adjacent grooves, is $D_r = R_{n+1} - R_n = m\lambda_{SPPC}$, where $\lambda_{SPPC}$ is the designed central SPP wavelength of the plasmonic demultiplexer. As a result, the $M$ point is the focal spot of the SPPs with a wavelength of $\lambda_{SPPC}$, and the focal spot shifts along the circle when the wavelength is changed. For different $m$ values, the coordinates of the groove centre $M$ are determined by $x_m = R\cos\theta\cos\theta$, $y_m = R\cos\theta\sin\theta$, where $\theta = \arcsin(m\lambda_{SPPC}/d)$. Therefore, the plasmonic demultiplexer can launch the input WDM signal to SPPs, split various wavelength components spatially, and then focus them into individual SPP waveguides.

   Three plasmonic demultiplexers with $D_r = 813.5$, 1,627, and 2,441 nm, which correspond to $m = 1$, 2 and 3 for $\lambda_{SPPC} = 813.5$ nm, were fabricated (details in the methods below). All of those structures had the same period $d = 4,068$ nm and total period number $N = 33$, which corresponds to a total length $L = 134$ $\mu$m along the $y$ axis. The radial width of the grooves $w$ = 407 nm and $R = 100$ $\mu$m. The scanning electron micrograph (SEM) of the plasmonic demultiplexer with $m = 3$ is shown in Figure 1b. In order to increase the SPP intensity, another two grooves were added to each period with a separation of 813.5 nm in the radial direction (Inset of Figure 1b). The projections of the three grooves in one period on the *y* axis are the same, so that they do not break the periodical nature along the *y* axis.

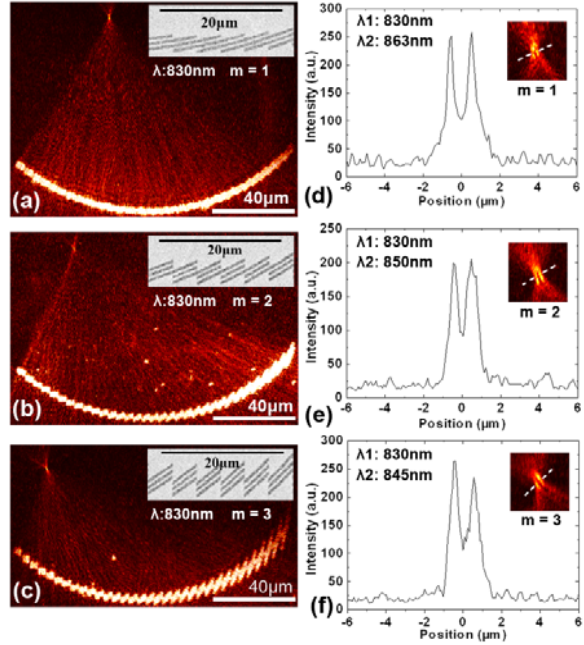

Figure 2. SPP images for *m* = 1 (a), 2 (b), and 3 (c). Insets are parts of the corresponding SEM images. Intensity profiles for *m* = 1 (d), 2 (e), and 3 (f) with wavelength differences 33, 20, and 15 nm, respectively, across the focus centres of the two focal spots (dashed line in the insets). Insets are the corresponding summed up SPP images around focal areas for two wavelengths.

In the experiment, a leakage radiation microscopy[22] (LRM) was used to detect the SPP images and the light polarization was adjusted to be parallel to the line *AM* (Figure 1a). The LRM SPP images of the plasmonic demultiplexers for *m* = 1, 2 and 3 are shown in Figure 2a, b, and c, respectively, and the parts of the corresponding SEMs of the demultiplexers are shown in the insets. The incident wavelength was 830 nm corresponding to a SPP wavelength of 813.5 nm. The directly transmitted light had been spatially filtered to make the SPP images clean.[22] The LRM images in the Fourier plane with and without filters are shown in Supporting Information Figure S1. In Figure 2, three focal spots are clearly seen and their locations agree with the discussion above. The distances between the three focal spots and the pole of the grating circle $A$ are 95, 88 and 76 $\mu$m, and the experimental transverse full widths at half maximum (FWHMs) of the focal spots are 645, 656, and 670 nm for $m$ = 1, 2 and 3, respectively. Then, the wavelength of the incident light was changed and the corresponding SPP images were recorded. In order to determine the resolutions of the plasmonic demultiplexers, two SPP images with different wavelength were summed up. Three results are shown in the insets of Figure 2d, e, and f with wavelength differences of 33, 20, and 15 nm for $m$ = 1, 2, and 3, respectively. The intensity distributions along the centers of the two focal spots (dashed lines in the insets) are depicted in Figure 2d, e, and f. The two focal spots with different wavelength are clearly separated for the three cases. As a result, the experimental resolutions $\Delta\lambda$ = 33, 20, and 15 nm are obtained for $m$ = 1, 2, and 3, respectively. It can be seen that a higher resolution is obtained with a higher order number $m$.

The distinctive characteristics of the plasmonic demultiplexers compared with optical demultiplexers include: intrinsically two-dimensional nature, the enhanced out-of-plane electronic field component, a higher propagation loss, and inhomogeneous exciting efficiency along the demultiplexer. However, the theoretical resolution of the plasmonic demultiplexers can be estimated as that of a conventional optical grating based on the Rayleigh criterion,[21] which is:

$$\Delta\lambda = n_{eff} \cdot \Delta\lambda_{SPP} = n_{eff} \cdot \lambda_{SPPC} /mN = \lambda_C /mN \qquad (1)$$

where $n_{eff}$ is the effective index of SPPs, and $\lambda_C$ is the corresponding central wavelength in vacuum. According to eq 1, the estimated theoretical wavelength resolutions of the above plasmonic demultiplexers are $\Delta\lambda = 25$, 13, and 8 nm for $m = 1$, 2, and 3, respectively, which are smaller than that of the experimental results. Besides the differences between the optical grating and the plasmonic demultiplexer, another reason is that the criterion used in the experiment is more critical than the Rayleigh criterion.

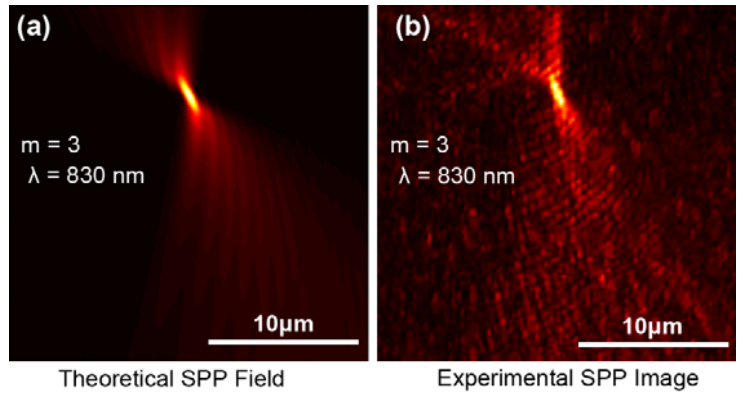

Figure 3. Comparison between the theoretical SPP image (a) and experimental result (b) for a incident wavelength of 830 nm and m = 3 around the focal area.

A more accurate calculation of the resolution was obtained by using the Huygens-Fresnel principle. Figure 3 (a) and (b) show the calculated SPP intensity and experimental result, respectively. It can be seen that the theoretical result agrees with the experimental one well, which validates our theoretical method. Two calculated SPP images with different wavelength were summed up to determine the resolution. By using this method, the theoretical resolutions of the above plasmonic demultiplexers were calculated to be 30, 15, and 10nm for $m = 1$, 2, and 3, respectively, based on the Rayleigh criterion (see Supporting Information Figure S2). The results are closer to the experimental results.

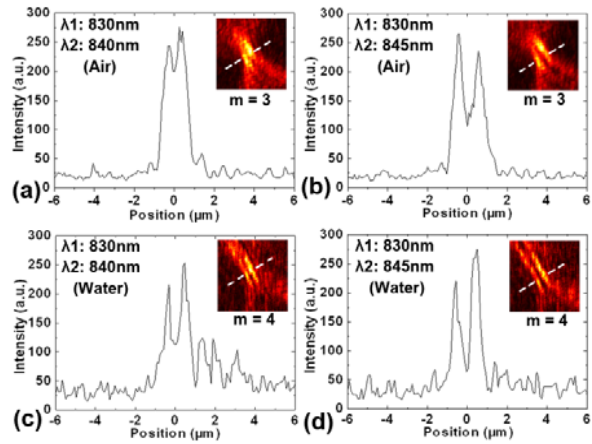

Figure 4. Comparison between the cases of air and water superstrates. Intensity profiles with wavelength differeces 10 (a) and 15 nm (b) for the case of air superstrate across the focus centres of the two focal spots (dashed line in the insets). Intensity profiles with wavelength differeces 10 (c) and 15 nm (d) for the case of water superstrate across the focus centres of the two focal spots (dashed line in the insets). Insets are the corresponding summed up SPP images around the focal areas.

According to eq 1, the resolution is determined by the product $mN$. In the case of $L = 134$ $\mu$m with an air superstrate, the maximum $mN$ was achieved when $m = 3$. Although the resolution can be improved by increasing the total length $L$, the size of the plasmonic demultiplexer is also increased. Here, we used a superstrate with a higher refractive index to decrease the SPP wavelength for obtaining a higher resolution. Water (refractive index of 1.33) was dropped on the gold film, and the SPP wavelength reduced to 602 nm for a vacuum wavelength of 830 nm. In this way, the maximum $mN$ was achieved with $m = 4$ while the total number of periods $N$ and the total length $L$ are the same as that in the case of air superstrate. The parameters of the fabricated plasmonic demultiplexer were as follows: $\lambda_{SPPC} = 602$ nm, $L = 134$ $\mu$m, $w = 301$ nm, $m = 4$ and $N = 33$. The total length was kept the same. Due to the shortening of the SPP wavelength, which results in a higher order number, the theoretical resolution according to eq 1 is improved to 6 nm. The SPP images with different incident wavelength were taken and summed up to compare with that of the case with an air superstrate and $m = 3$. The main results are shown in Figure 4. The intensity profiles were obtained along the centers of the two focal spots (dashed lines in the insets). It can be seen that the water superstrate evidently improves the resolution. A resolution of 10 nm was obtained, which is higher than the resolution with an air superstrate. When the wavelength difference was increased to 15 nm, the two spots were completely separated as shown in Figure 4d. At the same time, the FWHM of the focal spot decreased to 500 nm for the vacuum wavelength of 830 nm. In a practical plasmonic demultiplexer, a superstrate with a higher refractive index, such as InP that is widely used in the telecommunication system, can be used to obtain a higher resolution.

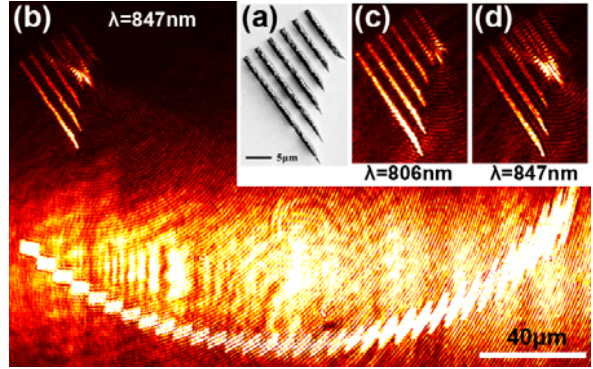

Figure 5. (a) SEM image of the strip waveguides. (b) SPP image for a incident wavelength of 847 nm. Enlarged SPP images around the waveguides for incident wavelengths of 806 (c) and 847 nm (d).

In a telecommunication system, the WDM signal that is spatially dispersed with respective to wavelength by the demultiplexer should be coupled to various waveguides. Metal strip waveguides have emerged for the advantages of confining light energy on subwavelength scales.[24-26] In the experiment, five strip waveguides with a width of 2 $\mu$m were fabricated using FIB and located on the focal circle of the above demultiplexer with an air superstrate and $m = 3$ (SEM image can be found in Supporting Information Figure S3). The detail SEM of the waveguides is shown in Figure 5a. The separation between two adjacent strip waveguides was 1 $\mu$m. The orientations of the strip waveguides were aligned to be parallel to the line *AM* (Figure 1a) and the entrance of each waveguide was located on the focal circle. Figure 5b shows the SPP image for a vacuum wavelength of 847 nm. Here, the directly transmitted light was not blocked for the sake of having a clear image of the waveguides (the image of SPP guiding in the waveguide with directly transmitted light blocked is shown in Supporting Information Figure S4). The fringes came from the interference between the SPPs and the directly transmitted light. The SPP images for the vacuum wavelengths of 806 and 847nm are shown in Figure 5c and d, respectively. It is clearly seen that the SPPs with two different wavelengths were fed into different waveguides and propagated along them. The fringes in the areas where gold was removed were caused by the interference between the directly transmitted light and the scattering light of SPPs by the gold remnant in these areas after FIB fabrication. The linear dispersion $\delta l$ of the plasmonic demultiplexer can be obtained by using the grating equation as:

$$\delta l = \frac{R \delta \lambda}{n_{eff} \sqrt{(\dfrac{d}{m})^2 - \lambda_{SPPC}^2}} \quad (2)$$

where $\delta \lambda$ is the wavelength difference of the incident light. In the experiment, $\delta l = 3 \ \mu$m, then $\delta \lambda = 33$ nm, which agrees with the experimental result 41 nm well.

In conclusion, we have proposed and experimentally demonstrated plasmonic demultiplexers by controlling the phase of SPPs excited at concentric grooves fabricated at dielectric/metal interface. Besides dispersing and multiple-channel SPP guiding, 3D propagating light mode to SPP modes coupling was also implemented simultaneously. Therefore, the plasmonic demultiplexers can act as converters to connect 3D conventional diffraction-limited photonic devices and 2D SPP devices in a WDM system. A resolution as high as 10 nm was obtained in the experiment and further improvement can be realized by using a superstrate with high refractive index. Due to the linear dispersion characteristic, the plasmonic demultiplexer can be used as a plasmonic spectroscopy or filter, and more applications, such as optical manipulation,[27] are also expected.

## METHODS

**Experiment.**

Gold film with thickness of 50 nm was thermally evaporated onto a 2 cm × 2 cm glass substrate. Then the plasmonic demultiplexers are fabricated on the gold film using a focused ion-beam (FIB) milling system. A laser beam from a tunable cw Ti:sapphire laser was focused by a cylindrical lens to illuminate the plasmonic demultiplexer. The polarization of the light source was tuned by a half-wave plate. A leakage radiation microscopy (LRM) was used to detect the SPPs, which consisted of an oil-immersion objective (100×, numerical aperture of 1.4), three lenses, and a spatial filter in front of Lens2 for spatially blocking the directly transmitted laser light through the gold film. The SPP intensity distribution was recorded by a charge-coupled-device (CCD) camera. The details of the LRM can be found in Ref. 22. (for the experiment setup, see the Supporting Information Figure S5).

**Theory.**

The theoretical calculation of SPP field was obtained by using the Huygens-Fresnel principle. Each point of the plasmonic demultiplexer was considered as a secondary SPP point source. The resulting SPP field can be obtained by adding up all the field of the SPP point source and taking into account its phase and amplitude. A complex phasor form of a SPP point source which located at the origin and $x$-polarized is[28]:

$$\vec{E}(\hat{\rho}, \hat{\varphi}, \hat{z})^{spp} = A(\hat{z} - \frac{i\alpha}{k_{spp}}\hat{\rho})H_1^{(1)}(k_{spp}\rho)\cos(\varphi)\exp(-\alpha z)\exp(-i\omega t) \quad (4)$$

where $A$ is a constant, $k_{\text{spp}}$ is the wave vector of SPP, and $H_1^{(1)}(k_{spp}\rho)$ is the $m = 1$ Hankel function. The propagation loss was considered in the calculation.

**RREFERENCES**